\documentclass[12pt]{article}
\usepackage{amssymb,cite,graphics}
\newcommand\blank[1]{}

\newcommand{\resection}[1]{\setcounter{equation}{0}\section{#1}}

\begin{document}
\begin{titlepage}
\vskip 0.5cm
\begin{flushright}
DTP-00/11 \\

{\tt hep-th/0002065} \\
\end{flushright}

\vskip 1.5cm
\begin{center}
{\Large{\bf
First order quantum corrections to the classical reflection
factor of the sinh-Gordon model
}}
\end{center}
\vskip 0.8cm
\centerline{A.
Chenaghlou$^{a,}$\footnote{{\tt alireza.chenaghlou@durham.ac.uk}}
and E. Corrigan$^{b,}$\footnote{\tt ec9@york.ac.uk}}
\vskip 0.9cm
\centerline{$^a$ \sl\small  Mathematical
Sciences,
University of Durham, Durham DH1 3LE, UK\,}
\centerline{$^b$ \sl\small Mathematics, University of York,
Heslington, York YO10 5DD, UK}
\vskip 2.25cm
\begin{abstract}
\noindent
The sinh-Gordon model is restricted to a half-line by
boundary conditions maintaining integrability. A
perturbative calculation of the reflection factor is
given to one loop order in the bulk coupling and to
first order in the difference of the two parameters
introduced at the boundary, providing a further
verification of Ghoshal's formula. The calculation is
consistent with a conjecture for the general dependence
of the reflection factor on the boundary parameters and
the bulk coupling.
\end{abstract}
\end{titlepage}
\setcounter{footnote}{0} 
\def\thefootnote{\fnsymbol{footnote}}

\section{Introduction}

Over the last few years, following the pioneering
ideas of Ghoshal and Zamolodchikov, and
others \cite{GZ,G,FKa,FKb,Sa},
much work
 has
been done to investigate
integrable quantum field
theory with a boundary. In particular, the affine Toda
field theories have offered a surprisingly rich structure
which is just beginning to be understood. The classical
affine Toda field theories are known to remain integrable
in the presence of certain (generally quite restricted) boundary
conditions confining them to a half-line, or to an
interval \cite{CDRS,CDR,BCDR,SS,FS,BCR,PZ,B,C1,D}.
The corresponding quantum field theories are hardly explored
although there has been progress in certain cases
associated with the $a_n^{(1)}$ class of models \cite{CDRS,Ga,DG,PB}.
The
first of these ($n=1$) is the sinh-Gordon model and, unlike all
the others has a set of integrable boundary conditions
depending on a pair of extra parameters, the so-called
boundary parameters. However, even in this case, it remains
to be seen precisely how the two boundary parameters
influence quantities of interest such as the reflection factor.

The sinh-Gordon model is related to the sine-Gordon
model and its reflection factors are related to the
reflection factors of the lightest breather in the
sine-Gordon model. Indeed, the results obtained by
Ghoshal and Zamolodchikov \cite{GZ,G} allow a determination
of the general form of the sinh-Gordon reflection factor
and this is a very useful piece of information. However,
apart from two special cases (Neumann and Dirichlet boundary
conditions) Ghoshal and Zamolodchikov's formulae fail
to provide a relationship between the reflection
factors and the boundary parameters themselves which
appear as the data in a Lagrangian formulation of the model.

Recently, it was noticed \cite{CD} that for certain ranges
of the boundary
parameters in the sinh-Gordon model there are real periodic
classical finite-energy solutions called boundary
breathers. The sinh-Gordon model has no real finite-energy
solutions at all on the whole line (other than $\phi=0$)
but, once there is a boundary singularities may be hidden behind
it allowing solutions restricted to a half-line to have
finite energy. The existence of periodic solutions allows
a semi-classical quantization determining the spectrum of
boundary bound states. Once the spectrum of bound state energies is
known it may be compared with a boundary bootstrap calculation
of the energies of the same states and hence may be used, in principle,
to find a relationship between the Lagrangian boundary parameters
and the data in Ghoshal's formula. In fact this is complicated
and was  carried out in \cite{CD} for the simpler situation
in which the
two boundary parameters are equal and the $\phi \rightarrow -\phi$
symmetry of the bulk theory is preserved. The corresponding
calculation in the more general case is not yet completed.
The semi-classical calculation may turn out to be exact
(as it did for the energy spectrum of the breathers
in the bulk sine-Gordon model itself; for a review see \cite{R})
but, failing a proof of exactness the results may be checked against
low-order perturbation theory \cite{K1,K2,C2,T}.

If the two boundary parameters are different and the $\phi\rightarrow -\phi$
symmetry is broken, the perturbation theory becomes substantially
more complicated: the lowest energy static background is no
longer the configuration $\phi =0$ and therefore the perturbation theory
must be developed within a non-trivial background, leading to
additional cubic and higher odd order couplings as well as a substantially
more intricate propagator. Nevertheless, as will be shown below,
provided calculations are restricted to first order in the
difference of the two
boundary parameters, some of the complications disappear and we are able
to calculate the correction to the reflection factor at one loop.
The result we obtain is consistent with a conjecture for
the relationship between the boundary parameters and Ghoshal's
formula in the more general setting. A full discussion, to all
orders in the difference of the boundary parameters, even at one loop
order in the bulk coupling, must be postponed.

\section{sinh-Gordon model}

The sinh-Gordon theory  corresponds to the affine Toda field
theory associated with
${a}_{1}^{(1)}$.
The bulk Lagrangian density of the theory is:
\begin{equation}\label{Lbulk}
\mathcal{L}=\frac{1}{2}\partial_{\mu}\phi\partial^{\mu}\phi
 -
V(\phi)
\end{equation}
where
\begin{equation}\label{Vbulk}
V(\phi)=\frac{2m^{2}}{\beta^{2}}\cosh(\beta \alpha \phi)
\end{equation}
The real constants $m$ and $\beta$ provide a mass scale
and a coupling constant, respectively, and it is customary
in affine Toda field theory to choose $\alpha = \sqrt{2}$.

Assuming two  sinh-Gordon particles of relative rapidity
$\Theta$
scatter from each other
elastically,  the bulk S-matrix characterizing this process
is
given by
\cite{FK,ZZ,AFZ}
\begin{equation}
S(\Theta)=-\frac{1}{(B)(2-B)}
\end{equation}
where
\begin{equation}\label{B}
B=\frac{1}{2\pi} \frac{\beta^{2}}{1+\beta^{2}/4\pi}
\end{equation}
and the symbol  $(\ )$ denotes the hyperbolic building
block:
\begin{equation}\label{block}
(x)=\frac{\sinh(\Theta/2+ \frac{i\pi
x}{4})}{\sinh(\Theta/2-\frac{i\pi
x}{4})}
\end{equation}
 The S-matrix is invariant under the
following weak-strong coupling
transformation

\begin{equation}
\beta \rightarrow 4\pi/\beta
\end{equation}
a property  known as  weak-strong coupling
duality.

A sinh-Gordon theory on the half-line
is described by the following  Lagrangian density:

\begin{equation}
\bar{\mathcal{L}}=\theta(-x) \mathcal{L} -\delta(x)
\mathcal{B},
\end{equation}
where $\mathcal{B}$ is regarded as a functional of the field only,
not its time
derivative. Moreover,
 the generic form of $\mathcal{B}$ is given by \cite{GZ}
\begin{equation}\label{Bboundary}
\mathcal{B}=\frac{m}{\beta^{2}}\left(\sigma_{0}
e^{-\frac{\beta}
{\sqrt{2}}\phi}+
\sigma_{1}e^{\frac{\beta}{\sqrt{2}}\phi} \right)
\end{equation}
Note, the coefficients $\sigma_{0}$ and $\sigma_{1}$ are a
pair of real
numbers, essentially free (but see \cite{CDR,FS}), which represent
the extra parameters permitted at the boundary $x=0$.

The equation of motion and
the boundary condition for the sinh-Gordon model on the
half-line become (after rescaling the mass scale to unity):
\begin{equation}
\partial^{2} \phi =-\frac{\sqrt{2}}{\beta}\left(
e^{\sqrt{2} \beta \phi}
-e^{-\sqrt{2} \beta \phi} \right) \hspace{.5in} \qquad  x<0
\end{equation}
\begin{equation}\label{Bcondition}
\frac{\partial \phi}{\partial x}=-\frac{\sqrt{2}}{\beta}
\left( \sigma_{1}
e^{\beta \phi/\sqrt{2}} - \sigma_{0} e^{- \beta
\phi/\sqrt{2}} \right)
\hspace{.5in} x=0
\end{equation}

\resection{Reflection factor}

 When a sinh-Gordon particle is moving towards the boundary
located at $x=0$ it will reflect elastically  from it
meaning that the in- and out- one particle states,
conveniently labelled by rapidity, will be related by
a reflection factor
\begin{equation}
|\theta>_{\rm out}=K(\theta)|-\theta>_{\rm in}.
\end{equation}

The general form of this reflection factor is known although
its detailed dependence
on the boundary parameters $\sigma_0$ and $\sigma_1$ appearing in
(\ref{Bboundary}) is
known only in
special cases. The sequence of arguments determining its
form is somewhat indirect, stemming from the work of Ghoshal and
Zamolodchikov
\cite{GZ} as follows. Solving the boundary Yang-Baxter equation,
and using general constraints implementing unitarity and a
form of crossing symmetry, it proved possible to calculate the
reflection factor for the sine-Gordon soliton. Since breathers
are soliton-anti-soliton bound states, a subsequent set of calculations
using the boundary bootstrap led Ghoshal \cite{G} to conjecture
reflection factors for each member of the full tower of breathers.
Finally,
the reflection factor for the lightest breather
 is supposed to be identical with that of the sinh-Gordon
particle provided
 the sine-Gordon  coupling $\beta$ is
replaced by $i\beta$.

Thus, suitably transformed in the manner described, Ghoshal's
formula for the sinh-Gordon reflection factor is given by:
 \begin{equation}\label{ghoshalformula}
K_{q}(\theta)=\frac{(1)(2-B/2)(1+B/2)}{(1-E(\sigma_{0},
\sigma_{1},\beta))
(1+E(\sigma_{0},\sigma_{1},\beta) )(1-F(\sigma_{0},
\sigma_{1},\beta))(1+F(\sigma_{0},\sigma_{1},\beta) )}
\end{equation}
Actually, Ghoshal's notation was a little different and made
use of two
 other quantities $\eta,\ \vartheta$ defined by
 $E=B\eta/\pi, F=iB\vartheta/\pi$.

 There are special cases in which $E$ and $F$ have been
 conjectured. For example,
 the Neumann boundary condition which is defined
 by
\begin{equation}
\frac{\partial\phi}{\partial x}=0 \hspace{.25in}\hbox{when}
\hspace{.25in}x=0,
\end{equation}
 has been argued by Ghoshal and Zamolodchikov to need a
 reflection factor containing
\begin{equation}\label{Eneumann}
F=0, \hspace{.25in} E=1-B/2.
\end{equation}

More recently \cite{CD},
 the
boundary
breather states of the sinh-Gordon model restricted to a
half-line were investigated and their energy spectrum
calculated in two
ways. First, by using the bootstrap equations, and then by
finding a set of periodic finite-energy solutions which
could be quantized using a WKB
approximation. Insisting that the two methods agreed
with each other
 provided strong evidence for a
relationship between
the boundary parameters, the bulk coupling constant, and the
 parameters
appearing in the reflection factor.
For technical reasons this work was restricted to the
case $\sigma_0=\sigma_1$ but, nevertheless, yielded
an expression for $E$, with $F=0$.
Specifically, setting
$\sigma_{0}=\sigma_{1}=\cos a\pi$ and restricting $a$
to the range $1>a>1/2$,
$E$ is given by,
\begin{equation}\label{Esimple}
E=2a(1-B/2).
\end{equation}
In the limit
$a\rightarrow 1/2$ from above, this is in agreement with
(\ref{Eneumann}).

Both of these conjectures are underpinned by low order
perturbation theory. Kim \cite{K1} has calculated the
one loop correction to the classical Neumann boundary
condition reflection factor and found agreement with
Ghoshal's formula to $O(\beta^2)$. On the other hand, a more
general calculation was carried out, also to one loop,
agreeing with the conjecture (\ref{Esimple}), and indeed,
preceding it \cite{C2}. In fact, the perturbative calculation in
\cite{C2} agrees with (\ref{Esimple}) for $a$ in the
range $1>a>0$.

However, when $\sigma_0\ne\sigma_1$ the situation becomes
much more complicated. Indeed, all that is known up to now
is the behaviour of $E$ and $F$ in the limit
$\beta\rightarrow 0$, deduced from a direct calculation of the
classical reflection factor in the general case \cite{CDR}.
To describe
this limit it is convenient to set $\sigma_0=\cos a_0\pi$
and $\sigma_1=\cos a_1\pi$, with $|a_i|\le 1,\ i=0,1$. Then,
\begin{equation}\label{EFclassical}
  E(0,\sigma_0,\sigma_1) =a_0+a_1, \qquad
  F(0,\sigma_0,\sigma_1)=a_0-a_1.
\end{equation}
and, the classical reflection factor itself is given in
term of the basic factors (\ref{block}) by
\begin{equation}\label{Kclassical}
K_{c}(\theta)=-
\frac{(1)^{2}}{(1-a_{0}-a_{1})(1+a_{0}+a_{1})
(1-a_{0}+a_{1})(1+a_{0}-a_{1})}.
\end{equation}

The expression (\ref{Kclassical}) is an essential ingredient of the
basic two-point function, or propagator, which takes the form \cite{C2},
\begin{eqnarray}\label{propagator}
G(x,t;x',t')&=&\int \int \frac{d\omega}{2\pi} \frac{dk}
{2\pi}\,
\frac{ie^{-i \omega (t-t')}}{(\omega^{2}- k^{2}-4 +i \rho)}\, f(-k,x^\prime)
\nonumber\\& &\hspace{40pt}\times \left(f(k,x) e^{ik(x-x')}
 +K_{c}f(-k,x)e^{-ik(x+x')}
\right)
\end{eqnarray}
where
\begin{equation}\label{fdef}
f(k,x)=\frac{ik-2 \coth 2(x-x_{0})}{ik+2}.
\end{equation}
The  parameter
$x_0$ enters the static background
\begin{equation}\label{background}
e^{\beta \phi_{0}/\sqrt{2}}=\frac{1+e^{2(x-x_{0})}}
{1-e^{2(x-x_{0})}}
\end{equation}
and determines the point at which the background becomes singular.
For this reason, it is crucial that $x_0\ge 0$. Actually, $x_0$
is determined by the boundary
condition (\ref{Bcondition}) and satisfies
\begin{equation}\label{xzero}
\coth x_{0}=\sqrt{\frac{1+\sigma_{0}}{1+\sigma_{1}}}.
\end{equation}
On the understanding that $\sigma_1\ge\sigma_0$, it
is enough to consider this situation since (\ref{Bcondition})
is symmetric under  $\phi\rightarrow -\phi$ and the
interchange of  $\sigma_0$ with $\sigma_1$.

In the expression (\ref{propagator}) for the propagator,
the classical reflection
factor appears
as the coefficient of the reflection part of
the free field
two-point function calculated within the classical
static background. To check Ghoshal's formula, the strategy
introduced  by Kim \cite{K1} and developed in \cite{C2} is to
calculate perturbative
corrections to the two-point function and then to identify corrections
to the classical reflection factor by picking out the coefficient of
$e^{-ik(x+x^\prime)}\ \hbox{as} \
 x , x' \rightarrow -\infty .$

\resection{First order quantum corrections to the reflection
factor}

Previously, as  noted above, the first order perturbation
calculation was achieved for the specially symmetric case
$\sigma_0=\sigma_1$.
In this section, this restriction will be relaxed and
enough of the ingredients of the perturbation expansion will
be calculated to allow a calculation (in principle) up to
$\beta^2$. We will discover that generally the theory
needs three- and four- point couplings which depend upon the
$x$-dependent static background.

Expanding the bulk potential (\ref{Vbulk})
around the
background
solution to the equation of motion, $\phi_{0}(x)$, we
derive the three-  and four-point couplings:

\begin{equation}
C_{\rm bulk}^{(3)}=\frac{2\sqrt{2}}{3}
\beta\sinh(\sqrt{2}\beta \phi_{0}),
\end{equation}
and
\begin{equation}
C_{\rm bulk}^{(4)}= \frac{1}{3}
\beta^{2}\cosh
(\sqrt{2}\beta \phi_{0}).
\end{equation}

On the other hand, the static background is
 (\ref{background}). So, after some manipulation,
we have
\begin{equation}
C_{\rm bulk}^{(3)}=\frac{4\sqrt{2}}{3} \beta\cosh 2(x-x_{0})
\left(
\coth^{2}2(x-x_{0}) -1\right),
\end{equation}
and
\begin{equation}
C_{\rm bulk}^{(4)}=\frac{1}{3} \beta^{2}\left( 2 \coth^{2}
2(x-x_{0})-1 \right).
\end{equation}

In the same manner, using (\ref{Bboundary}) we can derive the three-
and four-point
couplings
associated with the boundary term:
\begin{equation}
C_{\rm boundary}^{(3)}=\frac{\sqrt{2}
\beta}{12}\left(
\sigma_{1}e^{\beta \phi_{0}/\sqrt{2}}-\sigma_{0}e^{-\beta
\phi_{0}/\sqrt{2}}\right),
\end{equation}
and
\begin{equation}
C_{boundary}^{(4)}=\frac{\beta^{2}}{48}\left(\sigma_{1}
e^{\beta
\phi_{0}/\sqrt{2}}+\sigma_{0} e^{-\beta\phi_{0}/\sqrt{2}}
\right).
\end{equation}

For future reference, it will be useful to know these to first order
in the difference of the two boundary parameters
$\epsilon=\sigma_0-\sigma_1$:
\begin{equation}
C_{\rm bulk}^{(3)}=\frac{2\sqrt{2}}{3}\beta
\frac{\epsilon}{1+\sigma_{1}}e^{2x}+...
\end{equation}
\begin{equation}
C_{\rm bulk}^{(4)}=\frac{1}{3}\beta^{2}+...
\end{equation}
and similarly,
\begin{equation}
C_{\rm boundary}^{(3)}=\frac{\sqrt{2}\beta}{12}
\left(-\frac{\epsilon}{1+\sigma_{1}}\right)+...
\end{equation}
\begin{equation}
C_{\rm boundary}^{(4)}=\frac{\beta^{2}}{48}(2\sigma_{1}+
\epsilon)+...
\end{equation}

We shall also need the expressions for
$f(k,x)$ (\ref{fdef}) and the classical
reflection factor
$K_{c}$ (\ref{Kclassical}) to the same order in $\epsilon$
since both contribute to the propagator (\ref{propagator}).
In fact,
\begin{equation}
f(k,x)=1+O(\epsilon^{2})
\end{equation}
and the classical reflection factor (\ref{Kclassical}) reduces to
\begin{equation}
K_{c}=\frac{i k+2\sigma}{i k-2\sigma}+\frac{2i k}{(i
k-2\sigma)^{2}} \epsilon+ O(\epsilon^{2}).
\end{equation}
It is also convenient to write
\begin{equation}\label{Ksplit}
K_{c}=K_{0}+\epsilon K_{1},
\end{equation}
where $K_{0}$ is the classical reflection factor when the
two boundary
parameters  are equal.
\par
To calculate quantum corrections to the classical
reflection factor at
one loop order we use perturbative methods generalised
to the affine
Toda field theory on a half-line \cite{K1,K2,C2,T}.
(For earlier references on boundary perturbation theory in general
see;\cite{SZ,DD,BM} for affine Toda perturbation theory
see \cite{BCDS}, or the review \cite{C3}.) The $O(\beta^2)$ correction to
$K_0$ has been calculated before and the purpose of this article is to
calculate the corrections to $K_1$ to the same order.

The possible diagrams to $O(\beta^2)$ are:

\includegraphics
{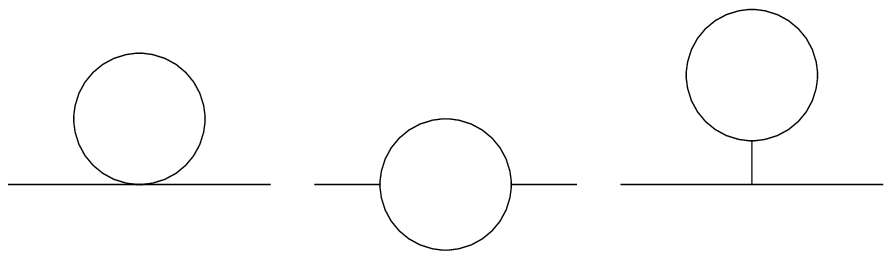}

\hspace{1.25in}I\hspace{1.15in} II\hspace{1.05in} III

\vspace{.25in}
\centerline{Figure 1: Three basic Feynman diagrams in one loop order.}

\vspace{.5in}
\noindent These will be computed in configuration space noting
that each vertex may either be situated at the boundary
or within the bulk. In effect, there are ten contributions which need
to be calculated.

However, there is a simplifying feature provided we are content to
work to first order in $\epsilon$. To recognise this
it is enough to note that because the three-point coupling
is already $O(\epsilon)$,
implying the type II and III
diagrams involve $\epsilon^2$,   only diagrams of type I need
concern us.

Thus, there are two contributions which need to be calculated
in order to be able to deduce the quantum corrections to the
classical reflection factor, both of type I. The first
 is directly
related to the boundary, when the interaction vertex
coincides with the
boundary, and it takes the form
\begin{equation}\label{Bbubble}
-\frac{i\beta^{2}}{4}(2\sigma_{1}+\epsilon)
\int_{-\infty}^{+\infty}dt''G(x,t;0,t'')G(0,t'';0,t''),
G(0,t'';x',t'),
\end{equation}
including the correct coupling constant and combinatorial factors.

The second contribution refers to the bulk potential
which means the
interaction vertex is located inside the bulk region $ x<0$.
 This contribution is given
by
\begin{equation}\label{Vbubble}
-4 i \beta^{2}\int_{-\infty}^{+\infty}dt''
\int_{-\infty}^{0}dx''
G(x,t;x'',t'')G(x'',t'';x'',t'')G(x'',t'';x',t').
\end{equation}
Again, the combinatorial factor has been
included.

Let us first calculate the boundary contribution
(\ref{Bbubble}). The loop propagator is given to
$O(\epsilon)$ by
\begin{equation}
G(0,t'';0,t'')=i \int \int
\frac{d\omega''}{2\pi}\frac{dk''}{2\pi}
P_0(\omega'',k'')
\left( 1+ \frac{i
k''+2\sigma_{1}}{i k''-2\sigma_{1}}+ \frac{2i k''
\epsilon}{(i k''-2\sigma_{1})^{2}} \right),
\end{equation}
where we have defined
\begin{equation}
P_0(\omega'',k'')
=\frac{1}{\omega''^{2}-k''^{2}-4+ i \rho}.
\end{equation}

Note, the  integral is clearly divergent but the
divergence is removed
by a renormalization of the boundary parameter. In effect,
rearranging the part of the integrand containing the
offending terms as follows,
\begin{equation}
1+\frac{i k''+ 2\sigma_{1}}{i
k''-2\sigma_{1}}=2+\frac{4\sigma}{i k''-2\sigma_{1}}
\end{equation}
and making  a minimal subtraction of the divergent portion,
(i.e. deleting the `2'),
renders the
integral finite.

The  part of the integral independent of $\epsilon$ has
been calculated before \cite{C2},
 therefore we may write,
\begin{equation}
G(0,t'';0,t'')=-\frac{a_{1}\cos a_{1} \pi}{\sin a_{1}\pi}
+ i \int \int
\frac{d\omega''}{2\pi} \frac {dk''}{2\pi}
P_0(\omega'',k'')
\frac{2i k'' \epsilon}{(i k''-2\sigma_{1})^{2}},
\end{equation}
and concentrate on the $O(\epsilon)$ part.

As far as the remaining integral is concerned, focusing on the energy
variable and
closing the integration contour in the upper half-plane,
we encounter a
simple pole at $\sqrt{k''^{2}+4}$. Therefore, we need
to evaluate
\begin{equation}
\frac{1}{2} \int \frac{dk''}{2\pi} \frac{1}{\sqrt{k''^{2}+
4}}\,
\frac{2 i k''}{(i k''-2\sigma_{1})^{2}} \epsilon\, .
\end{equation}
The $ k''$ integration may be performed by closing
the contour
in the upper (lower) half-plane depending on whether $\sigma_{1}>0$
($\sigma_{1}<0$). However, because of the branch points at $\pm 2i$
it is convenient to locate the associated branch cuts along
the imaginary axis from $\pm 2i$ to $\infty$.
Evaluating the integral and assembling the pieces, we
obtain
\begin{equation}
i \int \int \frac{d\omega''}{2\pi}\frac{dk''}{2\pi}
P_0(\omega'',k'')
\frac{2 i k''}{(i k''-2\sigma_{1})^{2}}
\epsilon=-\frac{\epsilon}{2}\frac{a_{1}}{\sin^{3}a_{1}\pi}
+\frac{\epsilon \cos a_{1}\pi}{2\pi}\frac{1}{\sin^{2}a_{1}
\pi}.
\end{equation}

At this stage, the contribution to
(\ref{Bbubble}) is
\begin{eqnarray}
& &-i\frac{\beta^{2}}{4}(2\sigma_{1}+\epsilon)
\left(-\frac{a_{1}\cos
a_{1}\pi}{\sin
a_{1}\pi}-\frac{\epsilon}{2}\frac{a_{1}}{\sin^{3}a_{1}\pi}+
\frac{\epsilon \cos a_{1}\pi}{2\pi}
\frac{1}{\sin^{2}a_{1}\pi} \right) \nonumber \\
&\times &i\int dt'' \int \int \frac{d\omega}{2\pi}\frac{dk}{2\pi}
e^{-i \omega(t-t'')}
P_0(\omega,k)\,
e^{-i
kx} \left( \frac{2i k}{i k-2\sigma_{1}}+\frac{2i k
\epsilon}{(i k-2\sigma_{1})^{2}} \right ) \nonumber \\
& \times&i\int \int \frac{d\omega'}{2\pi}\frac{dk'}{2\pi}
e^{-i \omega'(t''-t')}
P_0(\omega',k')\,
e^{-i k' x'}\left( \frac{2i k'}{i k'-2 \sigma_{1}}
+\frac{2i k' \epsilon}{(i k'-2\sigma_{1})^{2}} \right)
\end{eqnarray}
The integration over $ t''$ ensures energy
conservation at
the interaction vertex and creates a Dirac delta function
which immediately
removes one of the energy variables, for example $\omega'$.
The remaining integral
over the momenta $k$ and $k'$   can be performed by
completing the
contours in the upper half-plane and taking into
account the poles at
$k=k'=\sqrt{\omega^{2}-4}\equiv \hat{k}$. However, if
$\sigma_{1}>0$ it is
evident
that the expressions for $K_0$ and $K_1$
have no poles inside the contour. If $\sigma_{1}<0$,
there is an
additional pole but its contribution turns out to be
exponentially
decreasing in the asymptotic region $x,x' \rightarrow
-\infty$.

Finally, we obtain the boundary contribution (\ref{Bbubble})
in the form
\begin{eqnarray}\label{Bbubbleresult}
& &-i \frac{\beta^{2}}{4} \int \frac{d \omega}{2\pi}
e^{-i \omega
(t-t')} e^{-i \hat{k}(x+x')}
\left\{   \frac{2a_{1} \cos^{2}a_{1}\pi}{\sin a_{1}\pi}
\frac{1}{(i \hat{k}-2\sigma_{1})^{2}}\right. \nonumber \\
& &\hspace{.5in}+\left. \left(
\frac{a_{1}\cos a_{1}\pi}{\sin a_{1}\pi}+\frac{a_{1}
\cos a_{1}\pi}{\sin^{3}a_{1
}\pi}-\frac{\cos^{2}a_{1}\pi}{\pi \sin^{2}a_{1}\pi}
\right)
\frac{\epsilon}{(i \hat{k}-2\sigma_{1})^{2}}
\right.\nonumber\\
& &\hspace{2.5in}+\left.\frac{4a_{1} \cos^{2}a_{1}\pi}{\sin
a_{1}\pi}\frac{\epsilon}{(i \hat{k}
-2\sigma_{1})^{3}} \right\},
\end{eqnarray}
where $\hat{k}=2\sinh\theta$.

Next, we need to calculate the contribution (\ref{Vbubble})
which to $O(\epsilon)$ is:
\newpage
\begin{eqnarray}
& &-4i\beta^{2} \int dt'' \int_{-\infty}^{0} dx''
\int \int \frac{d\omega}{2\pi} \frac{dk}{2\pi}
e^{-i \omega (t-t'')}i P_0(\omega,k)\, \nonumber\\
& &\hspace{1in} \left( e^{-i k (x-x'')}\right.    
 \left.    +\frac{i k+2\sigma_{1}}{i
k-2\sigma_{1}}e^{-i k (x+x'')}+\frac{2i k \epsilon}{(i
k-2\sigma_{1})^{2}} e^{-i k(x+x'')}\right) \nonumber\\
& &\int \int
\frac{d\omega''}{2\pi}\frac{dk''}{2\pi}
i
P_0(\omega'',k'')
\left( 1+ \frac{i k''+2\sigma_{1}}{i
k''-2\sigma_{1}}
e^{-2i k'' x''}+\frac{2i k'' \epsilon}{(i
k''-2\sigma_{1})^{2}} e^{-2i k'' x''} \right) \nonumber \\
& &\int \int \frac{d\omega'}{2\pi}\frac{dk'}{2\pi}
i
P_0(\omega',k')\,
e^{-i
\omega'(t''-t')} \nonumber\\
& & \hspace{.75in}\left( e^{i k' (x''-x')}
\right.
+\left.\frac{i k'+2\sigma_{1}}
{i k'-2\sigma_{1}}
e^{-i k'(x''+x')}
 +\frac{2i k' \epsilon}{(i k'-2\sigma_{1})^{2}}
e^{-i k'(x''+x')} \right).
\end{eqnarray}
The integral over $ t''$ yields a delta function
which
replaces $\omega'$ by $\omega$.
Furthermore, to calculate
the integration over $ x''$, it is convenient to use the
following device
\begin{equation}
\int_{-\infty}^{0} dx'' e^{i k x'' +\tau x''}=\frac{-i}{k-i
\tau}
\end{equation}
where the small positive quantity $\tau$ will be taken to
zero
at the
final stage of the calculations.

The loop integral which
corresponds to the middle propagator of (\ref{Vbubble}),
is obviously
logarithmically
divergent and this divergence will be removed
by the infinite
renormalization of the mass parameter in the bulk
potential. So,
after making the subtraction and integrating
$x''$ and $\omega''$, and as before concentrating
on the $O(\epsilon)$ piece, we obtain the contribution
\begin{eqnarray}\label{bulk}
& &-\frac{i}{2}\int \int \int
\frac{d\omega}{2\pi}\frac{dk}{2\pi}\frac{dk'}{2\pi}
e^{-\imath\omega(t-t')}
e^{-i(kx+k'x')}
iP_0(\omega,k)
\, iP_0(\omega,k')
 \nonumber \\
& &\int \frac{dk''}{2\pi}\frac{1}{\sqrt{k''^{2}+4}}
\left\{ \frac{2i
k\epsilon}{(i k-2\sigma_{1})^{2}}\,\, \frac{i k''
+2\sigma_{1}}{i
k''-2\sigma_{1}}\left( \frac{1}{-k+k'-2k''-i \tau}
\right.\right.\nonumber\\
& &\hspace{3.2in}\left.\left.-\frac{1}{k+k'+2k''+i \tau}
 \,\,\frac{i k'+2\sigma_{1}}{ik'-2\sigma_{1}} \right)
\right.
\nonumber \\
& &\left.+ \frac{2i k' \epsilon}{(i k'-2\sigma_{1})^{2}}
 \,\,\frac{i k''+2\sigma_{1}}{i
k''-2\sigma_{1}}\left(\frac{1}{-k'+k-2k''-i\tau}-\frac{1}
{k'+k+2k''+i\tau} \,\, \frac{i k+2\sigma_{1}}
{i k-2\sigma_{1}}\right)
\right.
\nonumber
\\
& &\left.+\frac{2i k'' \epsilon}{(i
k''-2\sigma_{1})^{2}}\left(\frac{1}{k+k'-2k''-i
\tau}+\frac{1}{k-k'-2k''-i \tau}\,\,
\frac{i k'+2\sigma_{1}}{i
k'-2\sigma_{1}} \right.\right. \nonumber \\
& &\left. \left. -\frac{1}{k-k'+2k''+i \tau}
\,\,\frac{i
k+2\sigma_{1}}{i k-2\sigma_{1}}
\right.\right.\left.\left.-\frac{1}{k+k'+2k''+i
\tau}\,\, \frac{i k+2\sigma_{1}}{i k-2\sigma_{1}}\,\,
\frac{i
k'+2\sigma_{1}}{i k'-2\sigma_{1}} \right) \right\}.
\end{eqnarray}

In order to evaluate the integral over $k''$, we encounter the
following two
types:
\begin{equation}\label{typeone}
\int\frac{dk''}{2\pi}\,\,
\frac{1}{\sqrt{k''^{2}+4}}\,\,\left(\frac{i
k''+2\sigma_{1}}{i
k''-2\sigma_{1}}\right)\,\, \frac{1}{(k+k'-2k''-i \tau)}
\end{equation}
and
\begin{equation}\label{typetwo}
\int \frac{dk''}{2\pi}\,\,\frac{1}{\sqrt{k''^{2}+4}}\,\,
\frac{2i
k''
\epsilon}{(i k''-2\sigma_{1})^{2}}\,\,
\frac{1}{(k+k'-2k''-i
\tau)}.
\end{equation}
Both of these may be performed by closing an appropriate contour
in the
upper
half-plane ensuring that it runs around
the branch cut
located from $k''=2i$ to infinity along the
imaginary
axis. Note, if $\sigma_{1}>0$, then there is no pole
inside the contour; however, if $\sigma_{1}>0$, there is an
extra pole but
its residue
integrated over $k$ and $k'$  will give a contribution
vanishing
in the limit $x , x'$ $\rightarrow$ $-\infty$.

For example,  the integral (\ref{typetwo}) evaluates to
\begin{eqnarray}
& &\int \frac{dk''}{2\pi}\,\frac{1}{\sqrt{k''^{2}+4}}\,
\frac{2i
k''
\epsilon}{(i k''-2\sigma_{1})^{2}}\,
\frac{1}{(k+k'-2k''-i \tau)}
=\frac{\epsilon}{\pi}
\frac{k+k'}{(i k+i k'-4 \sigma_{1})^{2}}
\,\frac{a_{1}\pi}{\sin a_{1} \pi} \nonumber \\
& &\hspace{.5in}-\frac{\epsilon}{\pi} \frac{i
\sigma_{1}}{(4\sigma_{1}-i
k-i k')}\,\frac{1}{\sin^{2}a_{1}\pi}+
\frac{\epsilon}{\pi}\frac{i
\sigma_{1}^{2}}{(4\sigma_{1}-i k-i
k')}\,\frac{a_{1}\pi}{\sin^{3}a_{1}\pi} \nonumber \\
& &\hspace{.25in}+\frac{\epsilon}{\pi}
\frac{2i(k+k')}{(i k+i
k'-4\sigma_{1})^{2}}
\frac{1}{\sqrt{\frac{(k+k')^{2}}{4}+4}}
\ln{\left\{\frac{1+\frac{i(k+k')}{4}+\frac{i}{2}
\sqrt{\frac{(k+k')^{2}}{4}+4}}{1+\frac{i(k+k')}{4}-
\frac{i}{2}
\sqrt{\frac{(k+k')^{2}}{4}+4}}\right\}}
\end{eqnarray}

Let us divide the bulk contribution (\ref{bulk}) in two parts:
 $B_1$ containing integrals over $k''$ of type (\ref{typeone})
and $B_2$ whose $k''$ integrations are of
type (\ref{typetwo}).
For both it is necessary, after performing the integration over
$\omega''$, to do
the $k$ and $k'$ integrals  via
closing
contours in the upper half-plane to pick up the poles at
$k$ or $k'=\sqrt{\omega^{2}-4}=\hat{k} $. All
other pole contributions  lead to
exponentially damped
terms in the limit
$x,x'\rightarrow -\infty$.

After some manipulation, $B_{1}$ is found to be equal to
\begin{eqnarray}\label{Bone}
B_{1}&=& -2\beta^{2}\int \frac{d\omega}{2\pi}
e^{-i \omega(t-t')}
e^{-i \hat{k}(x+x')}\frac{1}{(2\hat{k})^{2}}
\frac{2i \hat{k}\epsilon}{(i
\hat{k}-2\sigma_{1})^{2}} \nonumber \\
& &\hspace{.5in}\left\{  -\frac{i}{4} \right.
\left. +\frac{i a_{1}}{\sin a_{1}\pi}\,
\frac{i
\hat{k}}{i
\hat{k}-2\sigma_{1}}
+\frac{1}{\pi} \frac{1}{\sqrt{\hat{k}^{2}+4}}\left(
\frac{i \pi}{2} -\theta\right) \right\}.
\end{eqnarray}

Notice that $B_{1}$, in the last term inside the braces,
depends on $\theta$  in a manner
which  potentially is very inconvenient
 for a comparison with
 Ghoshal's formula. Fortunately, this term will be
canceled by
a matching term in
$B_{2}$.

After somewhat lengthier calculations,  $B_{2}$  is
given by
\begin{eqnarray}\label{Btwo}
B_{2}&=&-2 \beta^{2} \int \frac{d\omega}{2\pi}
e^{-i \omega(t-t')}
e^{-i \hat{k}(x+x')} \frac{1}{(2\hat{k})^{2}}\left\{
\frac{-2i \epsilon}{(i
\hat{k}-2\sigma_{1})^{2}}\, \frac{\sigma_{1}^{2}}{\pi
\sin^{2}a_{1}\pi}
\right. \nonumber \\
& &\hspace{.25in}+\left.\frac{2i \epsilon }{(i
\hat{k}-2\sigma_{1})^{2}}\, \frac{a_{1}\sigma_{1}^{3}}
{\sin^{3}a_{1}\pi} + \frac{2i \epsilon \hat{k}}{(i
\hat{k}-2\sigma_{1})^{2}}
\, \frac{1}{\pi\sqrt{\hat{k}^{2}+4}}\  \theta \right.
\nonumber\\
& &\hspace{.5in}+\left.
\frac{i \hat{k}+2\sigma_{1}}{i \hat{k}-
2\sigma_{1}} \left( -\frac{i \epsilon}{2\pi
\sin^{2}a_{1}\pi}+
\frac{i \epsilon a_{1}\sigma_{1}}{2\sin^{3}a_{1}\pi}
\right) \right\}.
\end{eqnarray}

Assembling the $O(\epsilon)$ part of (\ref{Bbubbleresult}) with
(\ref{Bone}) and (\ref{Btwo}) we obtain,
\begin{eqnarray}
& &-\frac{i\beta^{2}\epsilon}{2}\int \frac{d\omega}{2\pi}
e^{-i\omega(t-t')}e^{-i \hat{k}(x+x')} \left\{
\frac{1}{(i
\hat{k}-2\sigma_{1})^{2}}\left(\frac{1}{2\pi}+
\frac{a_{1}\sigma_{1}}{2\sin a_{1}\pi}\right)
\right. \nonumber \\
& &\hspace{.1in}\left. + \frac{1}{(i
\hat{k}-2\sigma_{1})^{3}}\left(-2a_{1}\sin a_{1}\pi
\right)
+\frac{1}{(i \hat{k}-2\sigma_{1})^{2}}
\frac{2}{\hat{k}}\left(-\frac{i}{4}
+\frac{i}{2\sqrt{\hat{k}^{2}+4}} \right) \right\},
\end{eqnarray}
whence we can deduce the correction to the quantity $K_1$ in
(\ref{Ksplit}). Explicitly, we have,
\begin{eqnarray}\label{Kone}
\delta K_{1}&=& -i\beta^{2}\epsilon \hat{k}
 \left\{\frac{1}{(i
\hat{k}-2\sigma_{1})^{2}}\left(\frac{1}{2\pi}+
\frac{a_{1}\sigma_{1}}{2\sin a_{1}\pi}\right)+ \frac{1}{(i
\hat{k}-2\sigma_{1})^{3}}\left(-2a_{1}\sin
a_{1}\pi
\right)
\right. \nonumber \\
& &\hspace{1in}\left. +
\frac{1}{(i \hat{k}-2\sigma_{1})^{2}}
\frac{2}{\hat{k}}\left(-\frac{i}{4}
+\frac{i}{2\sqrt{\hat{k}^{2}+4}} \right) \right\}.
\end{eqnarray}
The correction to $K_0$ which was calculated before in \cite{C2}
is,
\begin{eqnarray}\label{Kzero}
\delta K_{0}&=&-\frac{i\beta^{2}}{8}K_{0}(\hat{k})
\sinh \theta
\left\{\left(\frac{1}{\cosh \theta+1}-\frac{1}{\cosh
\theta}\right)
\right.\nonumber\\
& &\hspace{.5in}\left. +2a_{1}\left(
\frac{1}{\cosh \theta-\sin
a_{1}\pi}-\frac{1}{\cosh \theta+\sin a_{1}\pi }\right)
\right\}.
\end{eqnarray}
This completes the collection of ingredients we need.

\resection{Comparison with Ghoshal's formula}

In this section, the corrections to the classical
reflection factor calculated above will be compared
with the formula of Ghoshal quoted in (\ref{ghoshalformula}).

Using (\ref{Ksplit}), the relative correction to the
classical reflection factor $K_{c}$ is given in terms
of the corrections $\delta K_0$ and $\delta K_1$ by
\begin{equation}
\frac{\delta K_{c}}{K_{c}}=K_{0}^{-1}\delta K_{0}
+\epsilon \left(K_{0}^{-1}
\delta
K_{1}-K_{1}
K_{0}^{-2} \delta K_{0}\right).
\end{equation}
Hence, using (\ref{Kone}) and (\ref{Kzero}) we have,
  \begin{eqnarray}\label{calculationfinished}
\frac{\delta
K_{c}}{K_{c}}&=&-\frac{i\beta^{2}}{8}\sinh \theta
\left\{\left(\frac{1}{\cosh \theta+1}-\frac{1}{\cosh
\theta}\right)
\right.\nonumber\\
& &\hspace{.5in} \left.+2a_{1}\left(
\frac{1}{\cosh \theta-\sin
a_{1}\pi}-\frac{1}{\cosh \theta+\sin a_{1}\pi }\right)
\right\}
\nonumber\\
& &+\frac{i\beta^{2}\epsilon \sinh\theta}
{8 \sin a_{1}\pi} \left\{
\frac{1}{\pi}\left(
\frac{1}{\cosh\theta-\sin a_{1}\pi}-
\frac{1}{\cosh\theta+\sin a_{1}\pi}\right)
\right.\nonumber\\
& &\hspace{.25in}+\left.a_{1}\cos a_{1}\pi\left(
\frac{1}{(\cosh\theta-\sin a_{1}\pi)^{2}}+
\frac{1}{(\cosh\theta+\sin a_{1}\pi)^{2}}\right)\right\}.
\end{eqnarray}

On the other hand, Ghoshal's formula (\ref{ghoshalformula}) for the
reflection
factor up to one loop
order is
given by:
\begin{equation}
K_{q}(\theta) \sim K_{c}(\theta) \left( 1-\frac{i
\beta^{2}}{8} \sinh \theta\
{\cal F}(\theta)
\right),
\end{equation}
where
\begin{eqnarray}\label{F}
{\cal F}(\theta)&=& \frac{1}{\cosh\theta+1} -
\frac{1}{\cosh\theta}\nonumber \\
& &+ \frac{e_{1}}{\cosh\theta+\sin(e_{0}\pi/2)}
-\frac{e_{1}}{\cosh\theta -
\sin(e_{0}\pi/2)}
\nonumber \\
& &+ \frac{f_{1}}{\cosh\theta+\sin(f_{0}\pi/2)}
-\frac{f_{1}}{\cosh\theta-\sin(f_{0}\pi/2)}.
\end{eqnarray}
In calculating (\ref{F}) we have made use of the expansions
of $E$ and $F$ to $O(\beta^2)$:
\begin{equation}
E\sim e_0 +e_1{\beta^2\over 4\pi} \qquad F\sim f_0 +
f_1{\beta^2\over4\pi},
\end{equation}
with
\begin{equation}
e_{0}=a_{0}+a_{1} \hspace{.5in} \hbox{and} \hspace{.5in}
f_{0}=a_{0}-a_{1}.
\end{equation}
Since $K_q=K_c+\delta K_c$, we deduce that
\begin{equation}
\frac{\delta
K_{c}}{K_{c}}=-\frac{i\beta^{2}}{8}\sinh \theta\  {\cal F}(\theta).
\end{equation}
Thence, expanding to $O(\epsilon)$, we find,
\begin{eqnarray}\label{expansion}
{\cal F}(\theta)&=&\left\{
\frac{1}{\cosh\theta+1}-\frac{1}{\cosh\theta}
+
\frac{e_{1}}{\cosh\theta+\sin a_{1}\pi}-
\frac{e_{1}}{\cosh\theta-\sin a_{1}\pi}
\right.\nonumber \\
& &\hspace{.1in}\left.+
\frac{e_{1}\epsilon \cos a_{1}\pi}{2\sin
a_{1}\pi}\left(
\frac{1}{(\cosh\theta+\sin a_{1}\pi)^{2}} +
\frac{1}{(\cosh\theta-\sin a_{1}\pi)^{2}}
\right)\right. \nonumber\\
& &\hspace{1.25in}\left.+\frac{\epsilon f_{1}}{\sin a_{1}\pi
\cosh^{2}\theta}\right\}.
\end{eqnarray}

Comparing (\ref{calculationfinished}) with (\ref{expansion})
we see a pleasing similarity. In fact the two formulae are identical,
to $O(\epsilon)$, provided we choose $e_1$ and $f_1$ suitably.
 In other
words,
we may deduce that
\begin{equation}
e_{1}=-2a_{1}+ \frac{\epsilon}{\pi \sin a_{1}\pi}\equiv
-(a_0+a_1) +O(\epsilon^2),
\end{equation}
and  $f_{1}$ is proportional to $\epsilon$. Unfortunately, the calculation
does not allow anything more detailed to be learned about $f_1$. To do better
would need a correction to the reflection factor to $O(\epsilon^2)$.

\resection{Discussion}

The purpose of this calculation was to test a little more deeply
the expression for the sinh-Gordon particle reflection factor given in
\cite{G} and to learn  additional information concerning its dependence on
the boundary parameters $\sigma_0$ and $\sigma_1$. The result
of the investigation is gratifying because it agrees with alternative
derivations of the reflection factor and it also agrees with the
following conjecture.
Everything we have learned so far is consistent
with quite simple expressions for $E$ and $F$:
\begin{equation}\label{conjecture}
E=(a_0+a_1)(1-B/2) \qquad F=(a_0-a_1)(1-B/2),
\end{equation}
where the coupling constant dependence enters via the expression for $B$
given in (\ref{B}). Similar expressions for these parameters
have been arrived at via other arguments by Zamolodchikov \cite{Z}.

If (\ref{conjecture}) is correct then the reflection factor
is invariant under the interchange $a_0\leftrightarrow a_1$. In
effect, this  invariance restores the
$\bf Z_2$ bulk symmetry which apparently was  broken by the boundary condition
and replaced by a symmetry under the simultaneous interchange of $\phi$ with
$-\phi$ and $a_0$ with $a_1$. The reflection factor is also invariant if
$a_0$ and/or $a_1$ is replaced by its negative, as it should be given the
definitions of $\sigma_0$ and $\sigma_1$. It is consistent with what is known at
the special value of the coupling constant, known as the `free-fermion'
point in the sine-Gordon model,
where $B=-2$ and the S-matrix is unity. There, the restrictions on the parameters
in the reflection factor can be solved exactly and are in agreement with
(\ref{conjecture}) \cite{C2}.

Note also, that with the expressions
(\ref{conjecture}) the reflection factor (\ref{ghoshalformula})
has a weak-strong coupling symmetry
matching the symmetry of the S-matrix under $\beta\rightarrow 4\pi/\beta$.
To see this, note that setting
\begin{equation}\label{duality}
(a_0^*,a_1^*,\beta^*)=\frac{4\pi}{\beta^2}(a_0,a_1,\beta)
\end{equation}
defines a new triple of coupling constants with the property that
\begin{equation}
K_q(\theta, a_0,a_1,\beta)=K_q(\theta, a_0^*,a_1^*,\beta^*).
\end{equation}

If (\ref{conjecture}) is correct, implying the duality symmetry (\ref{duality}),
then we are faced with other puzzles. For example, it is known that the
supersymmetric version of the sinh-Gordon model is only integrable when restricted
to a half-line with some very special boundary conditions (either $a_0=a_1=0$ or,
$a_0=a_1= 1$)
(see \cite{IOZ}), and this restriction would appear to be incompatible with
a weak-strong coupling symmetry without modifying (\ref{conjecture}).

It
is also known \cite{CDRS,BCDR} that the other affine Toda field theories
constructed from data
 in the $ade$
series, when restricted
to a half-line, allow only a finite number of possible boundary conditions.
In fact, the $a_1^{(1)}$ or sinh-Gordon model is apparently the only example
within this series which allows continuous boundary parameters. Expressions
for the associated reflection factors for the other models are largely unknown
but, it will be interesting to discover if they too can permit a duality
symmetry in the presence of a boundary which will match
the symmetry of their bulk S-matrices.

\resection{Acknowledgement}

One of us (AC) wishes to thank the Ministry of Culture and Higher Education
of Iran for financial support. The other (EC) thanks G.W. Delius and
Al.B. Zamolodchikov for discussions and the European Commission for
partial  support under a TMR grant, number FMRX-CT96-0012.


\end{document}